%% file: TripHill.tex
\documentclass[aps,prd,12pt,preprint,superscriptaddress,nofootinbib,floatfix]{revtex4-1}
\usepackage{latexsym}
\usepackage{amsfonts}
\usepackage[latin1]{inputenc}
\usepackage{subfig}
\usepackage{graphicx}
\usepackage{mathrsfs}
\usepackage{amssymb}
\usepackage{amsmath}
\usepackage{verbatim}
\usepackage[dvips]{color}
\newcommand{\slashD}{\slash\!\!\!\!\!\!\;D}

\begin{document}

\begin{flushright}
preprint FR-PHENO-2012-39\\
\today
\end{flushright}
\vspace*{1.0truecm}

\begin{center}
{\large\bf A singlet-triplet extension for the Higgs search at LEP and LHC}\\

\vspace*{1.0truecm}
{\large L.~Basso, O.~Fischer and J.J.~van~der~Bij}\\
\vspace*{0.5truecm}
{\it Physikalisches Institut, Albert-Ludwigs-Universit\"at Freiburg\\
D-79104 Freiburg, Germany}\\
\end{center}

\vspace*{1.0truecm}
\begin{center}
\begin{abstract}
\noindent 

We describe a simple extension of the standard model, containing a scalar singlet
and a triplet fermion. The model
can explain the possible enhancement in the decay $H \rightarrow \gamma \gamma$
at the LHC together with  the excess found in the Higgs boson search at LEP2.
The structure of the model is motivated by a recent argument, that was used to explain 
the number of fermion generations. For the sake of completenes we also considered the contributions
from higher multiplets.

\end{abstract}
\end{center}
\maketitle

\section{Introduction}
\label{sect:intro}
\input{intro}

\section{The model}
\label{sect:model}
\input{model}

\section{Results}
\label{sect:results}
\input{results}

\section{Conclusions}
\label{sect:conclusions}
\input{conclusions}

\section*{Acknowledgements}
\label{Sec:acknowledgements}
\input{acknowledgements.tex}

\bibliography{biblio}

\end{document}

%% file: intro.tex
The standard model (SM) describes particle physics in great detail. Nonetheless 
in the past most work has been based on the assumption, with varying motivations,
that the standard model must be incomplete and that new physics should be
just around the corner. With the new data from the LHC it appears reasonable to us
to question this assumption. The fact that the LHC has found no evidence for new physics
puts extensions in severe constraints. For instance the fact that LHCb finds full agreement
with standard model predictions basically rules out any new form
of flavour physics at accessible scales. Also direct searches have found no new physics, going
deep into the TeV scale. Therefore it appears to us that a new paradigm should be put into place.
Instead of asking what new physics there should be beyond the standard model, one should
look for the reason why the standard model is right and whether anything beyond the standard model
is possible at all. In the discussion of physics beyond the standard model, a lot of emphasis was put
on the so-called hierarchy problem, whereas some other questions tended to be ignored.
The most obvious of these questions are,
why there are precisely three generations of fermions and  why nature has chosen the gauge group that we see.
These questions have recently been addressed  by one of the authors~~\cite{vanderBij:2007fe,vanderBij:2010nu}. A condition was derived from a gravitational anomaly 
that appears to imply that, in the chiral sector, 
the standard model with three generations is the only possible low energy theory.
 Moreover, when grand unification is included, it was shown that Dirac triplets  could
be  expected in the low energy sector as well~\cite{vanderBij:2012ck}.

 The neutral component of such a triplet provides a WIMP-like cold dark matter candidate, the fermion number 
providing the needed unbroken symmetry. 
As we assume an exact fermion number conservation there is no connection with
the type-III seesaw~\cite{Foot:1988aq} in this case. Notice that triplet Dirac fermions alone cannot enhance the Higgs-to-diphoton channel. Such an enhancement requires an additional singlet scalar that mixes with the Higgs field.

Within the analysis of the LHC data the Higgs search plays an eminent role.
The data show a resonance at 126 GeV that is consistent with a standard model Higgs,
however the $H\rightarrow \gamma \gamma$ branching ratio appears to be enhanced.
Although the statistics is still limited, it is of interest to see whether the presence of extra triplet fermions could enhance this decay.

Moreover, in the search for the Higgs boson at LEP~\cite{Barate:2003sz} 
an excess at 98 GeV was found, that is compatible with the
presence of  a Higgs boson with roughly $10\%$ of the cross section of the standard model Higgs. We call such a boson a $10\%$ fractional Higgs boson. Such bosons generically appear when singlet scalars are present, that
 mix with the standard model Higgs (see \cite{vanderBij:2006ne} for a mini-review on singlet Higgses). 

If this excess is interpreted as a real signal, a reduction in the cross section for the 126 GeV Higgs of about $10\%$ is implied. In order to see whether the effects at the LHC and 
at LEP can be explained, we extend the standard model with an extra scalar singlet and a Dirac triplet. As a benchmark for singlet scalar extensions, we consider the Hill model~\cite{Hill:1987ea}. We will show that the effects can be easily described within this model. 

Many papers show effects on the diphoton decay of the Higgs boson, for a review see~\cite{Cacciapaglia:2009ky} and references therein. However only a few papers~\cite{Belanger:2012tt,Chang:2012ve,Drees:2012fb,Arbabifar:2012bd} consider a connection to the excess at LEP. The effect that we consider is motivated by an independent theoretical argument, namely unification within the constraint of the gravitational anomaly. Therefore we deem it useful to present our results, even though the calculations are straightforward.

This paper is organised as follows. In the next section we present the extension of the standard model with a singlet scalar.
For completeness sake we consider the coupling to a  generic $SU(2)$ multiplet Dirac fermion. In the subsequent  section we discuss the results on the $H \rightarrow \gamma \gamma$ decay, where we mainly focus on the triplet case. Finally, in the last section, we conclude.

%% file: model.tex
The starting point of our analysis is the Hill Higgs model~\cite{Hill:1987ea}.
In this model, the scalar Lagrangian reads
\begin{equation}
\mathscr{L} = -\frac{1}{2} \left( D_\mu \Phi \right) ^\dagger \left( D^\mu \Phi \right) - \frac{1}{2} \left(\partial _\mu H\right) ^2 - \frac{\lambda _1}{8} \left( \Phi^\dagger \Phi - v^2\right) ^2 - \frac{\lambda _2}{8} \left( 2f_2 H - \Phi^\dagger \Phi \right) ^2 \, ,
\end{equation}
where $\Phi$ is the Higgs doublet field and $H$ the scalar singlet Hill field. 
$H$ self-interaction terms are neglected. When no further particles are present this is consistent with renormalization
and allows for the determination of all parameters when two Higgs boson peaks are found.

The scalar potential is minimised for $\left< \Phi \right> = v = 246$ GeV and $\displaystyle \left< H \right> = \frac{v^2}{2 f_2} \equiv v_H$.
At the minimum, the $2$ {\it CP}-even scalars will mix as follows:
\begin{equation}
\left( \begin{array}{c} h_1\\ h_2 \end{array} \right) = \left( 
\begin{array}{cc} c_{\alpha} & s_{\alpha}\\ - s_{\alpha} & c_{\alpha} \end{array} \right) \, \left( \begin{array}{c} H \\ \phi \end{array} \right)\,,
\end{equation}
with $s_\alpha\,(c_\alpha)$ being the sine (cosine) of the mixing angle $\alpha$.
The mass eigenstates $h_{1(2)}$ couple to the SM particles with an overall $s_{\alpha}(c_{\alpha})$ prefactor and have masses
\begin{equation}\label{h_mas}
m_{h_{1,2}} = \frac{1}{2} \left( \lambda_2 f_2^2 + \lambda_3 v^2\right) \pm
\sqrt{\lambda_2^2 v^2 f_2^2 + \frac{1}{4}(\lambda_2 f_2^2 - \lambda_3 v^2)^2}\, ,
\end{equation}
where we have defined $\lambda_3 = \lambda_1 + \lambda_2$ and $m_{h_1} < m_{h_2}$. The mixing angle is given by
\begin{equation}\label{h_ang}
c_{\alpha}^2 = \frac{m^2_{h_2}-\lambda _2 f_2^2}{m^2_{h_2} - m^2_{h_1}}\, .
\end{equation}

To fit the data we invert eqs.~(\ref{h_mas})--(\ref{h_ang}) to express the parameters $f_2$, $\lambda_2$ and $\lambda_3$ in terms of the observable scalar masses and mixing angle:
\begin{eqnarray}\label{h_inv}
\lambda_2 &=&  \frac{s^2_\alpha c^2_\alpha (m^2_{h_2} - m^2_{h_1})^2}{v^2
(m^2_{h_1}+s^2_\alpha\,(m^2_{h_2} - m^2_{h_1}))}\, ,\\
\lambda_3 &=& \frac{m^2_{h_1}+c^2_\alpha\,(m^2_{h_2} - m^2_{h_1})}{v^2} \, ,\\
f_2 &=& v \frac{m^2_{h_1}+s^2_\alpha\,(m^2_{h_2} - m^2_{h_1})}{|s_\alpha| c_\alpha (m^2_{h_2} - m^2_{h_1})}\, .
\end{eqnarray}

In the following, we identify $h_2$ with the scalar boson recently observed at the LHC and we interpret the $2.3\sigma$ excess of Higgs-like events at $\sim 100$ GeV at LEP2~\cite{Barate:2003sz} as caused by the lighter scalar $h_1$ with couplings to the SM particles proportional to $s_\alpha$. Thus we set $m_{h_1}=98$ GeV, $m_{h_2}=126$ GeV and $0.05 < s_{\alpha}^2 < 0.15$, which defines 
$f_2 \simeq \mathcal{O}(1)$ TeV and $v_H \simeq \mathcal{O}(10)$ GeV. This shows that the presence of
two Higgs particles does not lead to any particular fine-tuning of parameters.

To fit the Higgs signals observed at the LHC, we extend the Hill Higgs model with fermion multiplet fields $\Delta$,  referred to as N-plets. 
Strictly speaking one should in this case extend the model with singlet-scalar self-couplings. However as we are not interested
in Higgs pair-production this plays no fundamental role in the following.
Results will be presented for only one N-plet, since the extension to more fermions is trivial.
We choose the latter to transform under the $3$ (triplet), $4$ (quadruplet), or $5$ (quintuplet) of the $SU(2)_L$ gauge group of the SM. The fermion N-plets transform as singlets under $SU(3)_C$ and have zero hypercharge, such that $\Delta\sim (1,N,0)$ with $N=3,4,5$. Higher multiplets are not considered here.

The N-plet Lagrangian reads
\begin{equation}
\mathscr{L}_{NP} = -\overline{\Delta}\, \slashD \Delta -y_{NP} H \overline{\Delta} \Delta - m_b \overline{\Delta} \Delta \, ,
\end{equation}
where a bare mass ($m_b$) term is allowed by the SM gauge symmetry. After spontaneous symmetry breaking the total N-plet mass is
\begin{equation}\label{N_mass}
m_{NP} = y_{NP} v_H + m_b \, .
\end{equation}

This shows that the mass of the N-plet is not proportional to its Yukawa coupling, unlike for the fermions of the standard model.
The Yukawa coupling and the mass of the N-plet can be considered as free parameters of the theory.  The contribution from the bare mass is vital to evade LEP limits, mainly the one coming from the $Z$-boson invisible decay width, constraining $m_{NP} > M_Z /2$. Further, we impose the constraint $m_{NP} > m_{h_2} /2 \sim 65$ GeV to prevent the Higgs from decaying into $2\Delta$, otherwise such decay channel would dominate the Higgs total width and therefore significantly reduce the branching ratios into SM particles. We have checked that such constraints are compatible with those coming from electroweak precision observables, in particular from the $S$ parameter. The contribution to the $T$ parameter vanishes.

The fit to data allows us to extract the scalar masses and the squared sine of the mixing angle. This leaves the possibility for $\sin{\alpha}$ to be either positive or negative. A negative sign would compensate the one from the fermion loop and thus lead to an enhanced diphoton signal for the Higgs boson, as observed at the LHC. Notice that a negative $\sin{\alpha}$ corresponds to a transformation such as $H \to -H$, which in turn can be seen as a change of sign of the Hill field's vev, or of the N-plet Yukawa coupling. Therefore putting the change of sign  in $\sin{\alpha}$ only, is not
a restriction on the theory.

An important ingredient of the mechanism proposed here is that the total N-plet mass does not come entirely from its interaction with the scalar fields, see eq.~(\ref{N_mass}). In other words, the coupling of the N-plet to the Higgs particle is {\it not} proportional to its mass. The former coupling is given by the Yukawa coupling constant $y_{NP}$, the latter mass is given by eq.~(\ref{N_mass}). In this way, we can separate the mass running in the loop from the strength of the interaction, i.e., light and strongly coupled particles are possible which can lead to a sizeable contribution to the Higgs-to-diphoton signal.

%% file: results.tex
\begin{figure}
\begin{center}
\includegraphics[width=0.7\textwidth,angle=-90]{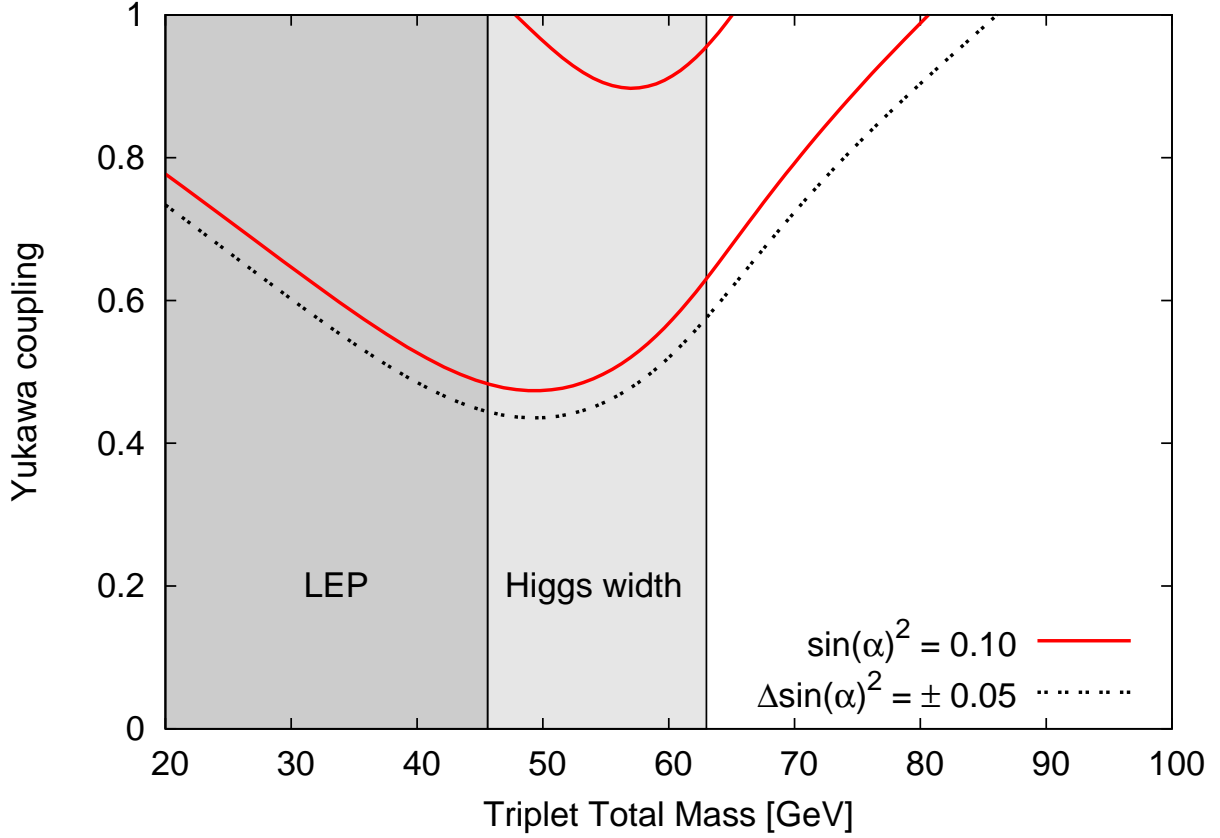}
\caption{Allowed 1-$\sigma$ range in the (mass--coupling) plane for one triplet to match the enhancement in the Higgs-to-diphoton signal. The red (solid) contour lines are for $s_\alpha^2=0.1$, while the black (dashed) line represents the variation of $s_\alpha^2$ with $\pm 0.05$, as from LEP~\cite{Barate:2003sz}. The allowed points are inside the lines. The shadings correspond to excluded regions.}
\label{fig_triplet}
\end{center}
\end{figure}

To start our discussion, a brief description of the decay width of the Higgs particle $h_i$ $(i=1,\,2)$ into two photons is presented. 
Ignoring the small effects of higher loops, this partial width can be written as
\begin{equation}
\Gamma_{\gamma \gamma}^i = \frac{G_F \alpha^2 m_i^3}{128 \sqrt{2} \pi^3} \left| A_1^i (\tau_W^i) + \sum_{\rm fermions} N_{c} Q_f^2 A_{1/2}^i (\tau_f^i) + \sum_{NP} Q_{NP}^2 A_{1/2}^i (\tau_{NP}^i) \right|^2\,,
\label{gammagamma}
\end{equation}
with the index $NP$ referring to the N-plets introduced before, $N_{c}$ the number of colours, $Q_x$ the electric charge, and $\tau$ given by
\begin{equation}
\tau_x^i = \frac{m_i^2}{4 m_x^2} \,.
\end{equation}
The relevant $A (\tau)$ functions, which depend on the spin of the particle in the loop and its couplings to the Higgs bosons, are~\cite{ Djouadi:2005gi}

\begin{eqnarray}
A_{1/2}^i (\tau) &=& \frac{2\,a_i}{\tau^2} \left( \tau + (\tau-1) f (\tau) \right)\,, \\
A_1^i (\tau) &=& - \frac{a_i}{\tau^2} \left( 2 \tau^2 + 3 \tau + 3 (2 \tau - 1) f (\tau) \right)\,,
\end{eqnarray}
with $a_{1(2)} = s_{\alpha}\, (c_{\alpha})$ due to the mixing in the scalar sector, and
\begin{eqnarray}
f (\tau) = \left\{ \begin{array}{lc}
\mbox{arcsin}^2 \sqrt{\tau}  & \tau \leq 1 \\
- \frac{1}{4} \left[ \log \frac{1+\sqrt{1-\tau^{-1}}}{1-\sqrt{1-\tau^{-1}}} - i \pi \right]^2 & \tau > 1 \,.\end{array} \right.
\end{eqnarray}
Given the scalar mixing and the fact that the N-plets couple to the Hill field only, we can define the N-plet amplitude as~\cite{Cacciapaglia:2009ky}
\begin{equation}
A_{NP}^i(\tau) = \sqrt{1-a^2_i} \, \frac{y_{NP}}{m_{NP}} \,\frac{2 \,v}{\tau^2} \left( \tau + (\tau-1) f (\tau) \right)\;,
\label{generalcoupling}
\end{equation}
to take into account the different Higgs coupling of the N-plets compared to that of the SM particles. It is clear that, as pointed out previously for the case of $h_2$, an enhancement is possible if we choose either the negative root (corresponding to a negative $s_\alpha$) or a negative Yukawa.
In the limit of large masses, we get a linear relation between $m_{NP}$ and the Yukawa coupling $y_{NP}$:
\begin{equation}\label{large_mass}
\frac{y_{NP}}{m_{NP}} = \frac{3}{4 \,s_{\alpha}\,v} \,\left( \sum_{NP} Q_{NP}^2\right)^{-1} \left[ A_1^2(\tau_W^2)+\sum_{\rm fermions} N_{c} Q_f^2 A_{1/2}^2 (\tau_{NP}^2)\right]\,\left(\sqrt{R^{\gamma\gamma}}-c_\alpha\right)\,,
\end{equation}
where we introduced the quantity
\begin{equation}
R^{\gamma\gamma}=\left. \frac{\sigma (pp\to h \to ff)_{obs}}{\sigma (pp\to h \to ff)_{SM}}\right|_{f=\gamma}\,,
\end{equation}
as measured at the LHC.

We compute the contribution of the N-plets to the processes $pp\to h_2 \to ff$ ($f=W,\,Z,\,b,\,\tau,\,\gamma$) and compare it with the experimental measurements, as listed in Table~\ref{table_observations}~\cite{ATLAS-CONF-2012-170,CMS-combination,CMS-combination_web}. 
In the no-$\gamma \gamma$ decays we averaged without taking into account correlations. 
This is not optimal, but it is the best that we can do with the published results. We urge the experimental collaborations to
perform this average, taking into account the systematic and statistical correlations in the data.
The average quoted is however sufficient to show that the LHC measurements are not in conflict with
the results of the Higgs search at LEP, which would imply a roughly 90\% signal strength at the LHC. For the following analysis
we only use the results from the $\gamma \gamma$ channel.

In the approximation that the production mechanism is dominated by the gluon-gluon fusion, we get a $c_{\alpha}$ modulation in the production amplitude (no extra contribution is present because our new fields are color singlets), and an analogous factor when the Higgs boson decays via tree-level processes, such as the decays into $WW$, $ZZ$, $b\overline{b}$, and $\tau^+\tau^-$. 
The case of the decay into photons is more complicated: being it a loop-induced process, its amplitude gets a $c_{\alpha}$ prefactor when $W$ bosons and tops are considered, and a $s_{\alpha}$ prefactor when the N-plets are mediating the process. Given that such channel is subleading in the evaluation of the Higgs total width, the latter also gets a $c_{\alpha}$ prefactor. 

Overall, the signal strength $R^{ff}$ (as compared to the SM) of the $pp\to h_2 \to ff$, where $f=W$, $Z$, $b$, and $\tau$, are all reduced by a $c_{\alpha}^2$ factor. 
The diphoton signal strength instead receives extra contributions from the N-plets, that results in a more complicated dependence from $s_{\alpha}$. For this to increase the diphoton signal strength, as argued before, either $\sin{\alpha}<0$ or $y_{NP}<0$ are required to compensate the negative sign coming from the fermionic loop.
 
\begin{table}
\centering
\begin{tabular}{|c||c|c||c|c||c|}
\hline
channel & CMS &  uncertainties & ATLAS &  uncertainties & average \\
\hline
$\gamma \gamma$ & 1.564 & $-$0.419,  +0.460 & 1.8 & $\pm$ 0.4 & 1.70 $\pm$ 0.30 \\ \hline
$ZZ$ & 0.807 & $-$0.280,  +0.349 & 1.0 & $\pm$ 0.4 & 0.97 $\pm$ 0.29 \\
$WW$ & 0.699 & $-$0.232,  +0.245 & 1.5 & $\pm$ 0.6 & 0.81 $\pm$ 0.23\\
$b\overline{b}$ & 1.075 & $-$0.566,  +0.593 & -0.4 & $\pm$ 1.0 & 0.69 $\pm$ 0.51\\
$\tau\tau$ & 0.875 & $-$0.484,  +0.508 & 0.8 & $\pm$ 0.7 & 0.85 $\pm$ 0.41\\
\hline
non-$\gamma$ av.& 0.78 & $\pm$ 0.18 & 0.97 & $\pm$ 0.29 & 0.83 $\pm$ 0.15 \\
\hline
\end{tabular}
\caption{Signal strengths for the Higgs boson observed at the LHC with respect to the SM expectation ($R$) for individual channels, and their average, at $M_h = 125$ GeV.}
\label{table_observations}
\end{table} 

Figure~\ref{fig_triplet} shows the range of masses and Yukawa couplings for the triplet to fit the observed diphoton signal strength. The portion of paramater space between the red (solid) contour lines matches the experimental observation for $s_\alpha^2=0.1$, while the black (dashed) contour lines denote the effect of varying $s^2_\alpha$ by $0.05$. We see that for allowed triplet masses ($m_{NP} \geq m_{h_2}/2$) and up to O(100) GeV, the observed enhancement of $R^{\gamma \gamma}$ can be matched, even with a single field (the case discussed here). The signal strengths of the other channels are consequently reduced by a factor $\cos{\alpha}^2 = 0.90 \pm 0.05$ as compared to the SM values, which is compatible with the values in Table~\ref{table_observations}. Given that the perturbative stability of the theory demands Yukawa couplings not bigger than $\mathcal{O}(1)$, an upper bound on the triplet mass of $\sim 90$ GeV for $s^2_\alpha = 0.1 \pm 0.05$ is obtained. This bound can be relaxed if more than one triplet is considered.

\begin{figure}
\begin{minipage}{0.49\textwidth}
\includegraphics[width=0.7\textwidth,angle=-90]{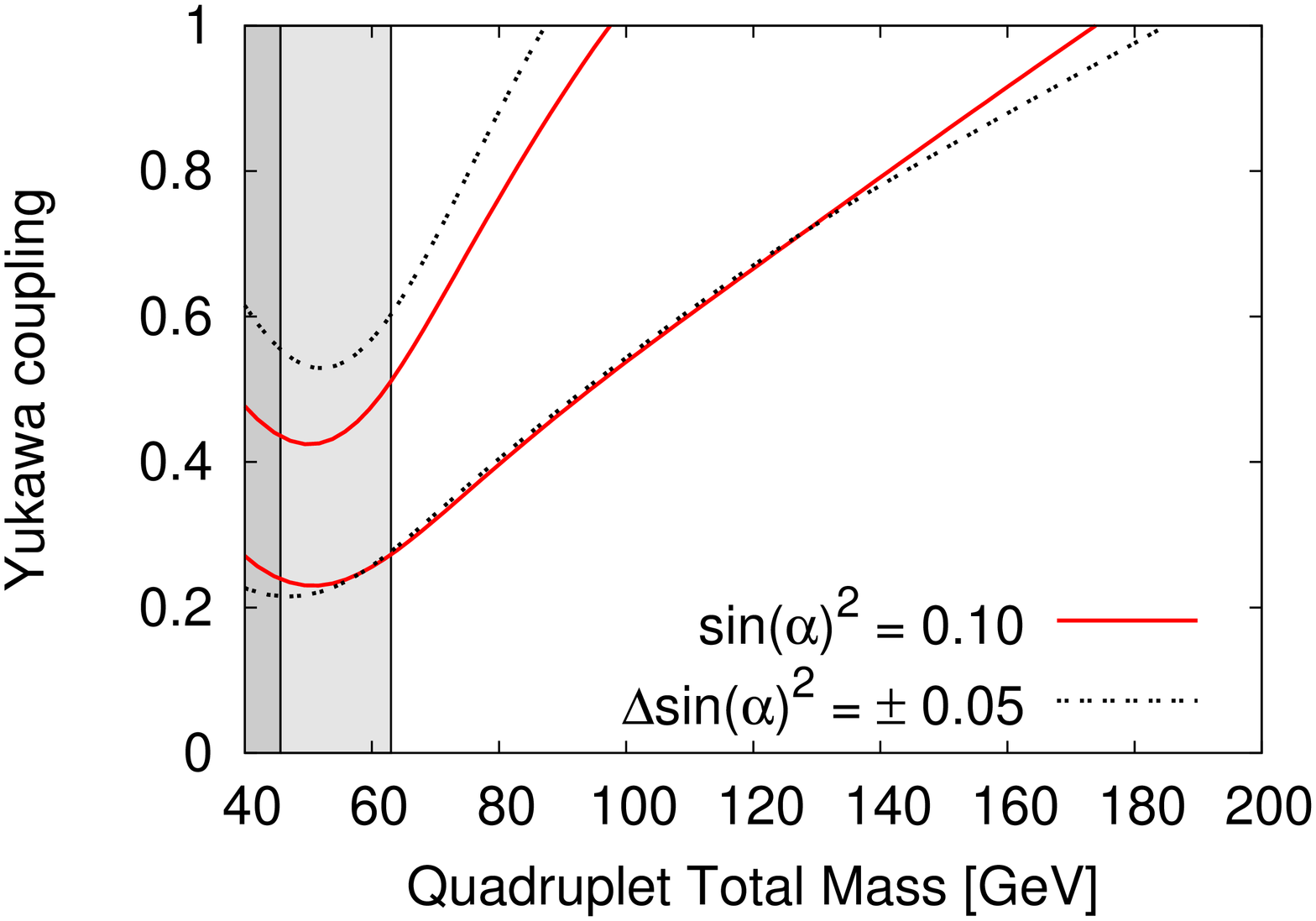}
\end{minipage}
\begin{minipage}{0.49\textwidth}
\includegraphics[width=0.7\textwidth,angle=-90]{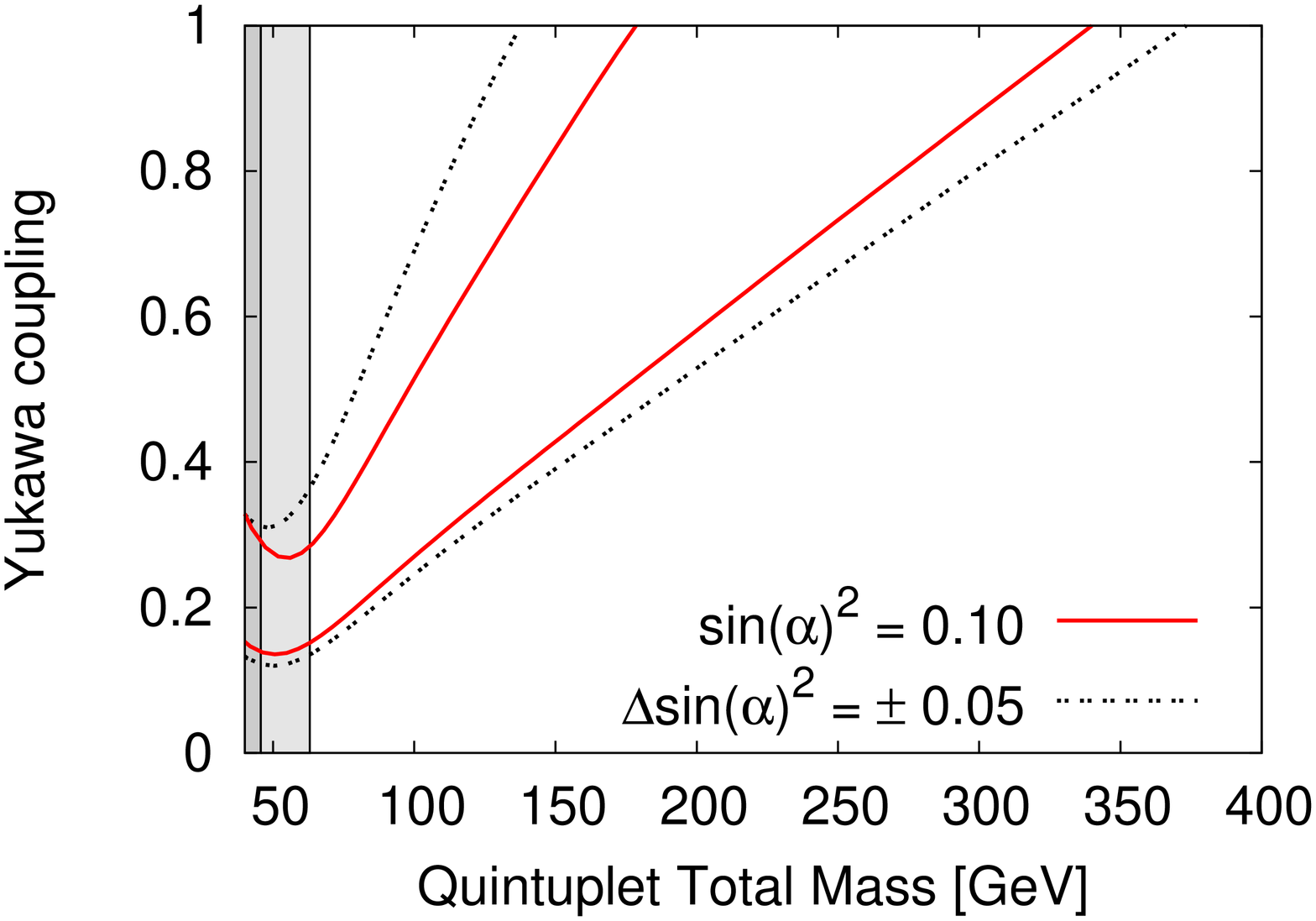}
\end{minipage}
\caption{Allowed 1-$\sigma$ range in the (mass--coupling) plane for (left) one quadruplet and for (right) one quintuplet, to match the enhancement in the Higgs to diphoton signal, for $s_\alpha^2=0.1$. The black (dashed) lines represent a variation of $s_\alpha^2$ of $\pm 0.05$, as from LEP~\cite{Barate:2003sz}.}
\label{fig_nplets}
\end{figure}

Figure~\ref{fig_nplets} illustrates the analogous situation for a quadruplet and a quintuplet field. Given that higher representations have a larger number of modes contributing to the diphoton channel
 (and some of these modes also have bigger electrical charges than the charged components in the triplet fields), their contribution is larger. This explains why bigger masses (per fixed Yukawa couplings) and/or smaller Yukawas (per fixed masses) are compatible with the experimental constraints. The demand of perturbative couplings again constrains the total N-tuple mass up to $\sim 200$ GeV for quadruplets and up to $\sim 400$ GeV for quintuplets for $s_\alpha$ in the considered range.

%% file: conclusions.tex
In this paper, we showed that both the observed enhancement of the
$H \rightarrow \gamma \gamma$ signal (if confirmed) and the excess of Higgs-like events at LEP2,
pointing to a $10\%$ fractional Higgs boson with a mass of $98$ GeV, can be explained in terms 
of a model with  an extra singlet scalar field and a Dirac triplet fermion. The latter is motivated by an argument used to explain the number of generations and the possibility of unification. Independent of this argument, the neutral component of a triplet is a candidate for at least part of the dark matter
in the universe~\cite{Cirelli:2005uq}.

The direct discovery of the fermion triplets at the LHC would be quite difficult. There is only a small mass splitting between the
charged and neutral components of the triplet. The signal would be a soft charged pion with missing energy~\cite{Cirelli:2005uq}. 
The same signal exists in  the case of scalar triplets.
In principle a discovery is not impossible, if the masses are smaller than about $150$ GeV~\cite{FileviezPerez:2008bj}.
The situation improves when one relaxes the Dirac condition on the fermions. If one splits the Dirac fermion
into two Majorana triplets one could make a larger mass difference, leading to more detectable signals.
Alternatively one could try to connect the triplets with the seesaw mechanism. In this case one violates lepton number
and loses the connection with dark matter. However this gives rise to more definite signals\cite{delAguila:2008hw}.

A concrete prediction of the model is that the production cross section of the $126$ GeV Higgs boson is reduced by about $10\%$ as compared to the standard model one. 
It is not clear whether the LHC can determine such a reduction, due to systematic errors, for instance PDF uncertainties, and theory errors. 

The detection of a signal of the $10\%$ fractional Higgs boson with a mass of $98$ GeV at the LHC seems to be a challenge at best, even for the high-luminosity option. To confirm or refute its
 presence, a new lepton collider is necessary.
Such a collider might also be needed to discover the triplet fermions.

However it is clear, that one first needs more statistics at the LHC to confirm the enhancement in the diphoton signal.

%% file: acknowledgements.tex
This work is supported by the 
Deutsche Forschungsgemeinschaft through the Research Training Group grant
GRK\,1102 \textit{Physics at Hadron Accelerators} and by the
Bundesministerium f\"ur Bildung und Forschung within the F\"orderschwerpunkt
\textit{Elementary Particle Physics}.